\documentclass[twocolumn]{aastex631}
\newcommand{\rev}[1]{#1}

\defcitealias{kobayashi21}{Paper~I}

\begin{document}

\title{
Rapid Formation of Gas-giant Planets via Collisional Coagulation from Dust Grains to Planetary Cores. II.  
Dependence on Pebble Bulk Density and Disk Temperature 
}

\shorttitle{
Collisional Growth from Non-Porous Dust Grains to Planets
}

\email{hkobayas@nagoya-u.jp}

\author{Hiroshi Kobayashi}

\affiliation{Department of Physics, Nagoya University, Nagoya, Aichi 464-8602, Japan}

\author{Hidekazu Tanaka}
\affiliation{Astronomical Institute, Tohoku University, Aramaki, Aoba-ku, Sendai 980-8578, Japan}

\begin{abstract}

\rev{Thanks to ``dust-to-planet'' simulations (DTPSs), which treat the collisional evolution directly from dust to giant-planet cores in a protoplanetary disk, we showed that giant-planet cores are formed in $\la 10\,$au in several $10^5$ years, because porous pebbles grow into planetesimals via collisions prior to drift in 10\,au \citep[][Paper I]{kobayashi21}.}
However, such porous pebbles are unlikely to reproduce the polarized millimeter wavelength light observed from protoplanetary disks. We thus investigate gas-giant core formation with non-porous pebbles via DTPSs. Even non-porous bodies can grow into planetesimals and massive cores to be gas giants are also formed in several $10^5$ years. The rapid core formation is mainly via the accretion of planetesimals produced by collisional coagulation of pebbles drifting from the outer disk. The formation mechanism is similar to the case with porous pebbles, while core formation occurs in a wider region (5\,--\,10\,au) than that with porous pebbles. 
Although pebble growth and core formation depends on the disk temperature, 
core formation is likely to occur with disk temperatures in typical optical thick disks around protostars. 
\end{abstract}

\section{Introduction}


Gas giant planets are commonly observed around stars \citep[$\sim 10\%$,][]{mayor11,fulton21}. 
The formation of gas giant planets is believed to be achieved from rapid gas accretion of a solid core with $\approx 10 M_\oplus$, where $M_\oplus$ is the Earth mass \citep[e.g.,][]{mizuno80,ikoma00}. The formation of such a massive core has a lot of difficulties. The formation timescale should be shorter than the disk lifetime (about a few million years). In addition, in a disk with gas densities corresponding to the minimum-mass solar nebula model, massive cores at 5\,au migrate inward in $\sim 10^5$ years \citep[e.g.,][]{ward97,tanaka02}. The core formation via planetesimal accretion is expected to be too slow to \rev{prevent the cores from migrating} onto the host star prior to the onset of rapid gas accretion \citep[e.g.,][]{kobayashi11,kobayashi18}. On the other hand, pebble accretion recently proposed \citep[e.g.,][]{lambrechts14} requires too much solid material to form cores with $\approx 10\,M_\oplus$ in usual protoplanetary disks with solid mass $\la 300\,M_\oplus$\citep{okamura21,kobayashi21}. 

\citet[hereafter Paper~I]{kobayashi21} carried out the simulation treating the collisional evolution from dust to \rev{giant-planet} cores in a whole disk, \rev{i.e. ``dust-to-planet'' simulations (DTPSs).} Collisional growth of dust aggregates forms pebbles. \rev{Further collisions between pebbles induce pebble growth} into planetesimals in $\la 10$\,au thanks to their low bulk density, while pebbles drift inward beyond $10$\,au. Once pebbles drift in $\la 10$\,au from the outer disk, they grow into planetesimals \rev{via collisions with other pebbles or planetesimals}. Cores are formed from planetesimals and further grow via accretion of new born planetesimals. The enhancement of planetesimals in $\la 10\,$au results in rapid growth of cores. Solid cores with $10\,M_\oplus$ are thus formed in several $10^5$ years. \rev{It should be emphasized that only collisional evolution is considered for the growth of bodies in DTPSs, where any models for planetesimal-formation instabilities are not applied.}

The direct growth of pebbles into planetesimals in inner disks is a key process
for the rapid formation of cores. The bulk density of pebbles plays an
important role for the planetesimal formation via collisional growth of
pebbles. 
\citetalias{kobayashi21}
assumed the bulk density of pebbles $\sim
10^{-3}\,{\rm g/cm}^3$. Such a low density is assumed according to the density evolution model based on the simulation results for sequential collisions with similar masses \citep{suyama12,okuzumi12}. 
However, the observation of the polarized
lights form protoplanetary disks are not explained with such a low bulk
density of large dust grains or pebbles \citep{tazaki19}. 
In addition, high mass ratio collisions of dust aggregates induces compaction of dust aggregates, resulting in the bulk density of $\sim 10^{-1}\,{\rm g/cm}^3$ \citep{tanaka23}.  
In this paper, we examine pebble bulk density dependence of 
the formation of gas giant cores via our DTPSs.
It will be seen that the disk temperature is another important foctor.
Thus we also examine the dependence on the disk temperature.
In Sections \ref{sc:disk_model} and \ref{sc:drift_growth}, we describe the disk model and analytical estimates
of time scales of dust drift and growth. 
In Section \ref{sc:DTPSmodel}, we briefly explain the model in DTPS.
In Section \ref{sc:result}, we examine the pebble bulk density and disk temperature dependence of formation of gas giant cores with DTPS. Our results are summarized \rev{and discussed} in Section \ref{sc:discussion}.

\section{Disk Model}
\label{sc:disk_model}

We consider a protoplanetary disk around a host star with mass $M_\star$. 
We apply the following gas and solid surface densities, $\Sigma_{\rm g}$ and $\Sigma_{\rm s}$, same as 
\citetalias{kobayashi21}. 
\begin{eqnarray}
 \Sigma_{\rm g} &=& \Sigma_{\rm g,1} (r/1\,{\rm au})^{-1},\label{eq:sigmag} \\
 \Sigma_{\rm s} &=& \Sigma_{\rm s,1} (r/1\,{\rm au})^{-1},
\end{eqnarray}
where $\Sigma_{\rm g,1} = 480 \, {\rm g/cm}^2$ and $\Sigma_{\rm s,1} = 8.5 \,{\rm g/cm}^2$ are the gas and solid
surface densities at $1\,{\rm au}$, respectively, and $r$ is the distance from the host star. The outer edge radius determines the disk mass. According to the assumption in 
\citetalias{kobayashi21}, 
we set the outer edge radius of 108\,au, which corresponds to the total solid mass of $\approx 220\,M_\oplus$. 

In 
\citetalias{kobayashi21}, 
we applied the temperature model similar to \citet{hayashi85}, although the applied model has lower temperature than the original model because of the radially optical thickness. 
The disk temperature is important to discuss pebble drift (see Equation~\ref{eq:tdrift}). 
However, the model applied in 
\citetalias{kobayashi21}
was too crude to discuss the drift accurately. 
We \rev{here} apply the temperature model determined by the balance between stellar irradiation and thermal emission in an optically thick disk. 
The temperature at the disk midplane is given by \citep{kusaka70} 
\begin{equation}
 T = T_1 \left(\frac{r}{1\,{\rm au}}\right)^{-3/7},\label{eq:temperature} 
\end{equation}
where $T_1$ is the temperature at 1\,au, given by
\begin{eqnarray}
 T_1 &=& \left(\frac{L_\star}{28 \pi \sigma_{\rm SB}}\right)^{2/7} \left(\frac{2k_{\rm B}}{m_{\rm mol} G M_\star}\right)^{1/7} (1\,{\rm au})^{-3/7}, \\
&=& 1.0 \times 10^2 \left(\frac{L_\star}{L_\sun}\right)^{2/7} \left(\frac{M_\star}{M_\sun}\right)^{-1/7}\,{\rm K},\label{eq:t1} 
\end{eqnarray}
where $L_\star$ is the luminosity of the host star, $L_\sun$ is the solar luminosity, $M_\sun$ is the solar mass, $\sigma_{\rm SB}$ is the Stefan-Boltzmann constant, $k_{\rm B}$ is the Boltzmann constant, $G$ is the gravitational constant, and $m_{\rm mol}$ is the mean weight of gas molecules. 

Note that the mid-plane temperature given by Equation~(\ref{eq:temperature}) is almost same as that by the two-layer model \citep{chiang01} or for the disk structured by stellar winds \citep{yun13}. In addition, the accretion heating is ignored in the temperature model, which may be important even at 1\,au if disks have massive accretion. Although we normalize the temperature at 1\,au with $T_1$, 
the temperatures are mainly determined by Equation~(\ref{eq:temperature}) in 
outer disks at $\ga 5$\,au to which we are concerned for planet formation.

\section{Analytic Estimate}
\label{sc:drift_growth}

The pebble sized bodies have ${\mathrm St} \sim 1$, where ${\mathrm St}$ is the stopping time multiplied with the Keplarian frequency $\Omega = \sqrt{G M_\star/r^3}$. 
The behavior of drifting pebbles with ${\mathrm St} \approx 1$ is important for the formation of gas-giant cores via collisional coagulation 
\citepalias{kobayashi21}. 
The drift timescale of the pebbles is estimated to \citep{adachi76}. 
\begin{eqnarray}
 \tau_{\rm drift} &=& \frac{14 m_{\rm mol} r^2 \Omega}{19 k_{\rm B} T} 
\frac{1}{{\mathrm St}}. 
\label{eq:tdrift} 
\end{eqnarray}
Equation~(\ref{eq:tdrift}) shows the drift timescale is inversely proportional to $T$. Therefore the disk temperature is important to discuss the behavior of pebbles as mentioned in \S~\ref{sc:disk_model}. 

The growth of pebbles occurs via collisions between similar sized pebbles. 
The collisional growth timescale of pebbles, $\tau_{\rm grow} \equiv s/\dot s$, is estimated to  \citep[e.g.,][]{ohashi21}
\begin{equation}
\tau_{\rm grow} = 
4
\sqrt{\frac{\pi}{3}}
\frac{\rho_{\rm b} s}{\Sigma_{\rm s} \Omega} \frac{\phi}{{\mathrm St}},\label{eq:growth} 
\end{equation}
where $\rho_{\rm b}$ and $s$ are the bulk density and radius of pebbles, respectively and 
\rev{$\phi$ is the correction factor for the mass spectrum of pebbles. 
In the derivation of Equation~(\ref{eq:growth_time}), the relative velocities between pebbles and the pebble scale height are assumed to be determined by turbulent stirring and are given by $(2 \alpha_{\rm D} {\mathrm St})^{1/2} c_{\rm s}$ and $(2 \alpha_{\rm D}/3 {\mathrm St} )^{1/2} c_{\rm s}/\Omega$, respectively, where $c_{\rm s}$ is the sound speed and $\alpha_{\rm D}$ is the dimensionless turbulent strength.}

If pebbles are not so fluffy $\rho_{\rm b} \gg 10^{-3}\,{\rm g/cm}^3$, 
pebbles with ${\mathrm St} \la 1$ are sized smaller than the \rev{gas molecular} mean free path\rev{, $\lambda_{\rm mfp}$}. Thus, for $s < 9 \lambda_{\rm mfp}/4$, $s = 2 \Sigma_{\rm g} {\mathrm St} / \pi \rho_{\rm b}$. 
\rev{In addition, the mass distribution of pebbles is self-similar, which gives $\phi \approx 1.2$. }
Inserting the relation into Equation~(\ref{eq:growth}), we have 
\begin{equation}
 \tau_{\rm grow} = 
4
\sqrt{\frac{\pi}{3}}
\frac{\Sigma_{\rm g} \phi}{\Sigma_{\rm s} \Omega}.\label{eq:growth_time} 
\end{equation}
Using Equations~(\ref{eq:tdrift}) and (\ref{eq:growth_time}) with the disk model described in \S\ref{sc:disk_model}, 
we estimate the ratio of the timescales, 
\begin{equation}
 \frac{\tau_{\rm grow}}{\tau_{\rm drift}} \approx 0.56 \, {\mathrm St} 
  \left(\frac{r}{10\, {\rm au}}\right)^{4/7} 
  \left(\frac{T_1}{100\,{\rm K}}\right)
  \left(\frac{M_\star}{M_\sun}\right)^{-1}, \label{eq:timescale_ratio} 
\end{equation}
\rev{where we used $\phi = 1.2$. } 
In the inner disk with low $T_1$, we have $\tau_{\rm grow} / \tau_{\rm drift} \la 1$ so that pebbles grow into planetesimals with ${\mathrm St} > 1$ \rev{unless collisional fragmentation disturbs. According to impact simulations for icy dust aggregates, collisional fragmentation of pebbles is negligible \citep[e.g.,][see also the next section]{hasegawa23}}. Therefore, the temperature is very important for the growth of pebbles.  

On the other hand, if pebbles are very fluffy, pebbles with ${\mathrm St} \la 1$ are larger than the mean free path. The ratio of the timescales additionally have a factor of $(9 \lambda_{\rm mfp} / 4 s)$. If $\rho_{\rm b} \la 10^{-3} \, {\rm g/cm}^{3}$ in the inner disk of $r \la 10\,$au, the factor is much smaller than the unity because of large $s$. Then $\tau_{\rm growth} /\tau_{\rm drift} \la 1$  so that the growth of pebbles is achieved 
\citep[\citetalias{kobayashi21}]{okuzumi12}. 
However, such low density pebbles are unlikely to explain the millimeter-wave polarization of protoplanetary disks \citep{tazaki19}. 

\section{Model for DTPS}
\label{sc:DTPSmodel}

We carry out simulations for the collisional evolution of bodies from dust to planets (``Dust-to-planet'' simulation; DTPS) developed in 
\citetalias{kobayashi21}. 
We mainly follow the method proposed in 
\citetalias{kobayashi21}. 
We here briefly explain the model of DTPS. 

In DTPS, we calculate the mass spectrum of bodies from sub-micron sized grains to gas-giant core sized bodies in annuli into which the disk is divied, which evolves due to collisions between bodies and their radial drifts \rev{caused by gas drag or by disk-planet interactions (type I migration). The mass spectrum is expressed by mass bins with the mass ratio between adjacent bins of 1.05. }
The collisional rate is determined by the number densities of bodies, the relative velocities, and the collisional cross-sections. The number densities are determined by the mass spectrum and the scale height of bodies. 
For ${\mathrm St} \la 1$, bodies are controlled by the gas motion so strongly that the relative velocities and the scale heights are determined by the Brownian motion, the settling, the radial drift, and/or the turbulence. On the other hand, large bodies with ${\mathrm St} \ga 1$ have random velocities and scale heights \rev{that are determined by orbital eccentricities and inclinations 
evolving by the collision-less encounters (orbital interactions)}, the collisional damping, the turbulence stirring, and the gas drag, whose evolution is calculated simultaneously with collisional evolution. 
In addition, we take into account the bulk densities of bodies, the gravitational focusing, the gas damping during encounters, and planetary atmospheres to determine the collisional cross-sections. \rev{DTPSs can treat the formation of gas-giant cores at the onset of runaway gas accretion because we do not consider the gas accretion of cores due to the collapse of hydro-static atmospheres.} Our calculations can reproduce $N$-body simulations for planetesimal accretion \citep{kobayashi+10}. Furthermore, we performed the collisional evolution of bodies from sub-micron sized grains to gas-giant cores 
\citepalias{kobayashi21}. 

We ignore collisional fragmentation according to the discussion in 
\citetalias{kobayashi21}. 
\rev{We still lack realistic laboratory studies of high velocity collisions between icy aggregates in conditions equivalent to protoplanetary disks. However, collisional }simulations of icy dust aggregates with sub-micron sized monomers show collisional fragmentation is insignificant unless the relative velocities are larger than 50\,--\,80\,${\rm m/s}$ \citep{wada13}. Recent optical and near-infrared polarimetric observations of protoplanetary disks are consistent with 
dust aggregates composed of sub-micron sized monomers \citep{tazaki22,tazaki23}.  
In addition, even relatively high velocity collisions result in mass transfer between target aggregates to projectile aggregates, which are quit different from collisional fragmentation \citep{hasegawa21}. Therefore, significant fragmentation does not disturb collisional growth. We will address collisional fragmentation according to the collisional outcome model given by \citet{hasegawa23} to investigate the evolution of protoplanetary disks observed by multiple wavelengths. 
\rev{On the other hand, collisional fragmentation of planetesimals is induced by the formation of gas-giant cores \citep[e.g.,][]{kobayashi+10,kobayashi11}. However, we totally ignore the collisional fragmentation for simplicity. }

We below introduce the difference from the model in 
\citetalias{kobayashi21}. 

\subsection{Bulk Density}

The collisional growth of dust grains produces the fractal dust
aggregates with masses $m$ and radii $s$. 
The bulk densities $\rho_{\rm b} = (3 m / 4 \pi
s^3)$ decrease due to early collisional growth and then increase due to the compaction by ram pressure and self-gravity. \rev{Radioactive heating can also induce the compaction for planetesimals. We do not consider the effect of radioactive heating because significant compaction is important for pebble-sized bodies. }

In \citetalias{kobayashi21},
we modeled $\rho_{\rm b}$ 
as
\begin{equation}
 \rho_{\rm b} = \left[\rho_{\rm mat}^{-1} + (\rho_{\rm s} + \rho_{\rm m} + \rho_{\rm l})^{-1}\right]^{-1}, 
\end{equation}
where 
$\rho_{\rm mat} = 1.4 \,{\rm g/cm}^3$ is the material density, corresponding to the density of compact bodies or monomer grains in dust aggregates, 
$\rho_{\rm s}$, $\rho_{\rm m}$, and $\rho_{\rm l}$ determine 
the bulk density for small, intermediate, and large bodies, respectively. 
For small dust, collisional evolution decreases the bulk density, while the bulk density increases for large dust by self-gravity. They are modeled as \citep{okuzumi12,kataoka13}
\begin{eqnarray}
 \rho_{\rm s} &=& \rho_{\rm mat} \left(\frac{m}{m_{\rm mon}}\right)^{-0.58},\label{eq:rho_s}
\\
\rho_{\rm l} &=& \left(\frac{256 \pi G^3 \rho_{\rm mat}^9 s_{\rm mon}^9  m^{2}}{81 E_{\rm roll}^3 }\right)^{1/5},\label{eq:rho_l} 
\end{eqnarray}
where $s_{\rm mon} = 0.1\,\micron$ is the monomer radius, $m_{\rm mon} = 4 \pi \rho_{\rm mat} s_{\rm mon}^3/3$ is the monomer mass, and $E_{\rm roll}=4.74\times 10^{-9}\,{\rm erg}$ is the rolling energy between monomer grains. 

The minimum value of bulk densities corresponds to $\rho_{\rm m}$, the bulk density of intermediate bodies. Therefore, $\rho_{\rm m}$ is a parameter in this paper. 
\citetalias{kobayashi21}
applied $\rho_{\rm m} = 10^{-3} \, {\rm g/cm}^3$ according to the sequential impact simulation of similar-mass dust aggregates \citep{suyama12}. However, \citet{tanaka23} showed sequential collisions with high mass ratios result in higher bulk densities. We estimate the resultant pebble bulk densities of $\approx 0.5\,{\rm g/cm}^3$ in the given disk under the assumption that all collisions have high mass ratios. This estimate is the high-end bulk density resulting from sequential collisions because similar mass collisions reduce the bulk densities. Therefore, we newly apply $\rho_{\rm m} = 0.1 \, {\rm g/cm}^3$ to investigate planet formation from non-porous pebbles, which are likely for protoplanetary disk observations. 

\begin{figure*}[hbt]
\plottwo{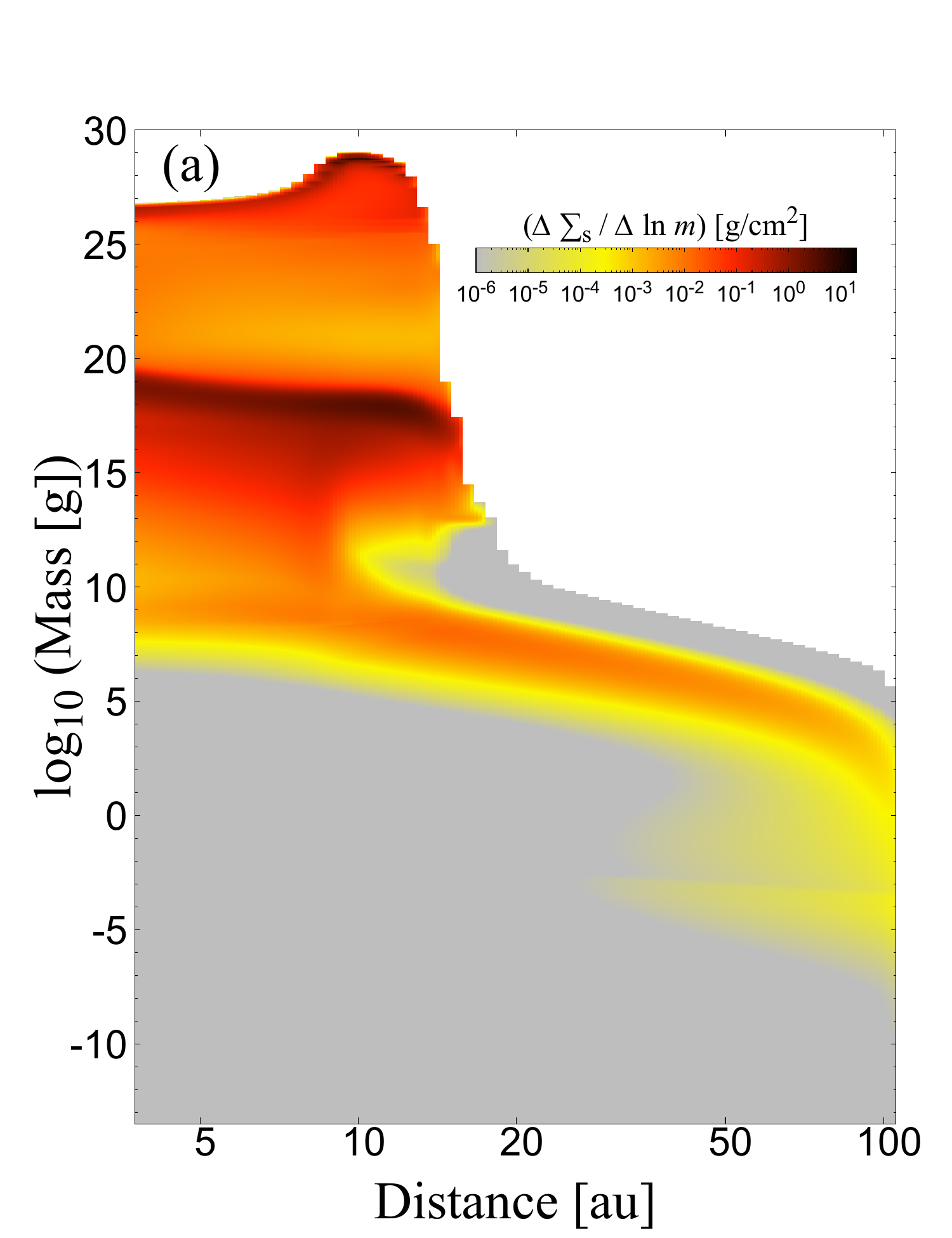}{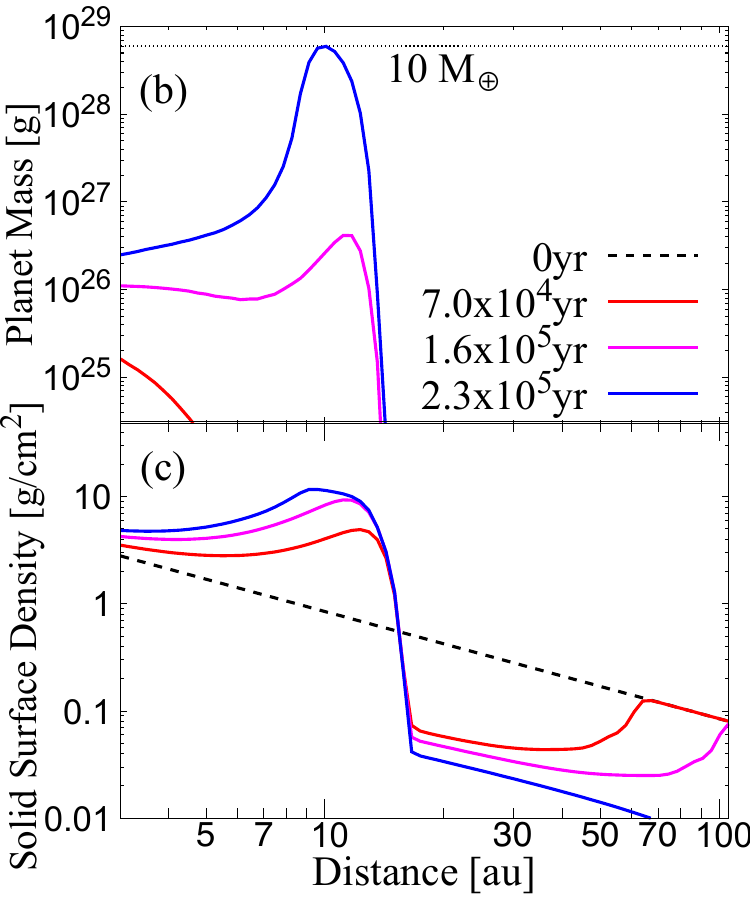}
\figcaption{Results of DTPS with $\rho_{\rm m} = 10^{-3} \,{\rm g/cm}^{3}$ and $T_1=100$\,K. Left: (a) Solid surface Density, $\Delta \Sigma_{\rm s} / \Delta \ln m$ at $t=2.3 \times 10^5$ years, as a function of the mass of bodies and the distance from the host star. The values of the solid surface density are shown in the color bar.
Right: Protoplanet mass (b) and total solid surface density (c), as a function of the distance from the host star. 
} \label{fig:dist_d1e-3}
\end{figure*}

\section{Results of DTPS}
\label{sc:result}
\subsection{Dependence on Bulk Density}

We here carry out the DTPSs with $T_1 = 100$\,K and $M_\star = M_\sun$. Figure \ref{fig:dist_d1e-3} shows the mass distribution of bodies at $2.3 \times 10^5$ years as a result of the DTPS with $\rho_{\rm m} = 10^{-3} \,{\rm g/cm}^3$. 
The mass distribution is expressed by the surface density of mass $m$ at $r$, $\Delta \Sigma_{\rm s} / \Delta \ln m$, defined as 
\begin{equation}
 \frac{\Delta \Sigma_{\rm s}}{\Delta \ln m} \equiv m^2 n_{\rm s}, 
\end{equation}
where $n_{\rm s} dm$ is the surface number density of bodies with masses between $m$ and $m+dm$ at the distance $r$. 
As shown in 
\citetalias{kobayashi21}, 
the collisional growth of porous dust aggregates directly produce planetesimals with $m \ga 10^{\rm 15}$\,g in the inner disk $\la 10$\,au. The runaway growth of planetesimals then forms protoplanets, which leads to the high $\Delta \Sigma_{\rm s}/\Delta \ln m$ zone around $m\sim 10^{19}$\,g inside $\la 10$\,au \citep{kobayashi16}.\footnote{\rev{The runaway growth is stalled by turbulence for small planetesimals. The growth of planetesmals induces their runaway growth due to strong mutual gravitational interactions. 
Collisional growth of largest planetesimals increases the peak masses in surface densities prior to runaway growth, while the runaway growth of planetesimals forms the small number of large planetesmals without significant growth of peak masses \citep[e.g.,][]{kobayashi+10}. 
The onset of the runaway growth of planetesimals determines the peak masses in surface densities \citep{kobayashi16}. The mutual collisions between planetesimals however increase the peak masses slowly \citep{kobayashi+10}. Therefore, the peak masses are slightly larger in the inner disk.}} Protoplanets grow further via the accretion of planetesimals \citep{kobayashi+10,kobayashi11}. In the outer disk ($\ga 10$\,au), dust aggregates grow up to pebble sizes with $m \sim 10^{4}$\,--\,$10^{9}$\,g, which drift inward by gas drag. The drifting pebbles are seen at the high $\Delta \Sigma_{\rm s} / \Delta \ln m$ zone at $m\approx 10^4$\,--\,$10^9$\,g in Figure~\ref{fig:dist_d1e-3}. 
Once pebbles drift in the inner disk ($\la 10$\,au), the collisional coagulation of drifting pebbles produces small planetesimals. The enhancement of small planetesimals formed from pebble growth induces the rapid growth of protoplanets. In $t \approx 2 \times 10^5 $ years, a core with $10 M_\oplus$ is formed. The disk parameter assumed above results in a core formation at $\approx 10$\,au, while the core formation occurs at 6\,--\,7\,au in the parameter given in 
\citetalias{kobayashi21}. 

The evolution of the protoplanet mass and the total surface mass density is shown in Figure~\ref{fig:dist_d1e-3}b,c. The mass of a protoplanet becomes $\approx 10\,M_\oplus$ at $2.3 \times 10^5$ years, which subsequently forms a gas giant planet via rapid gas accretion. The rapid formation of gas-giant cores is largely caused by the enhancement of the solid surface density in the inner disk.

\begin{figure*}[htb]
\plottwo{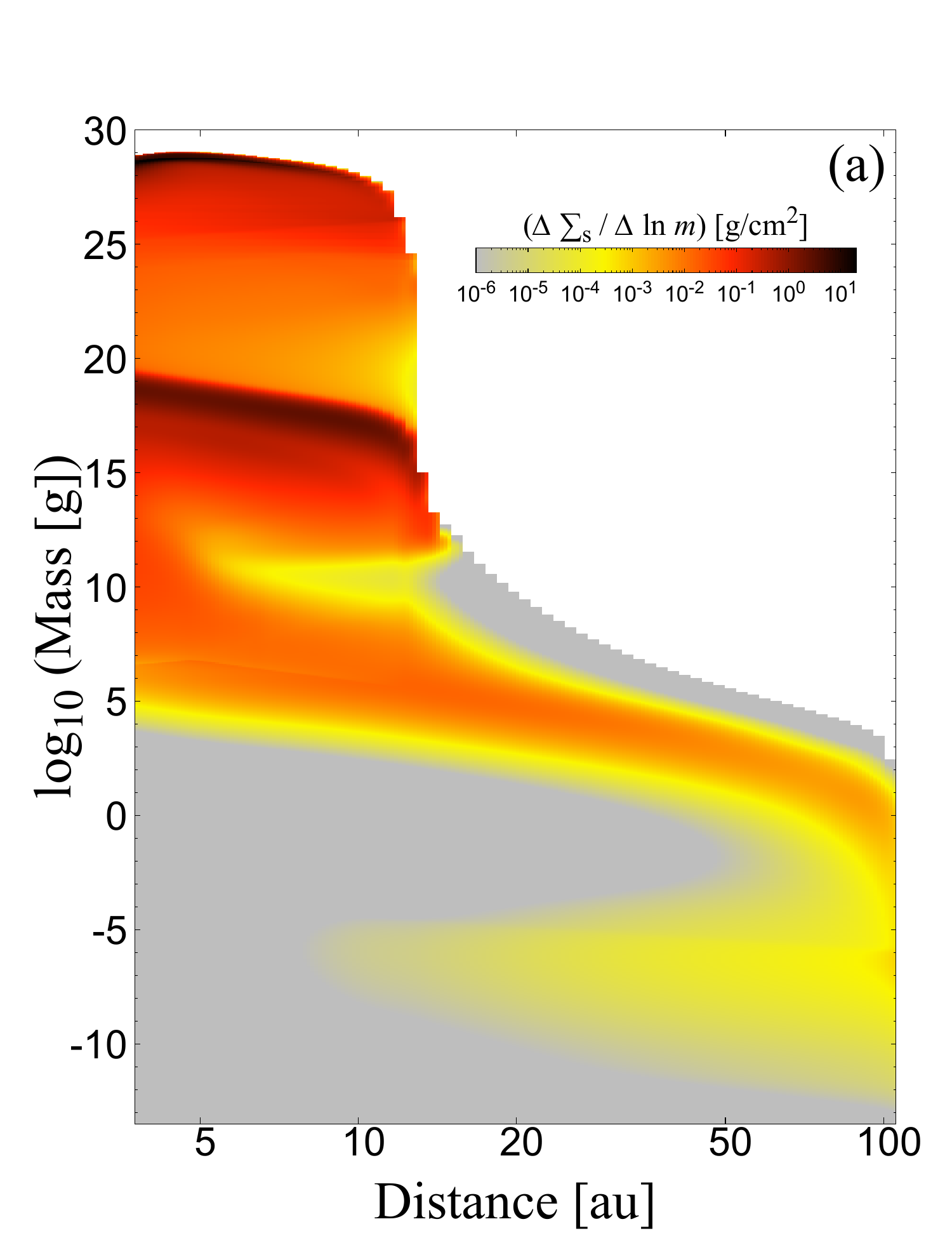}{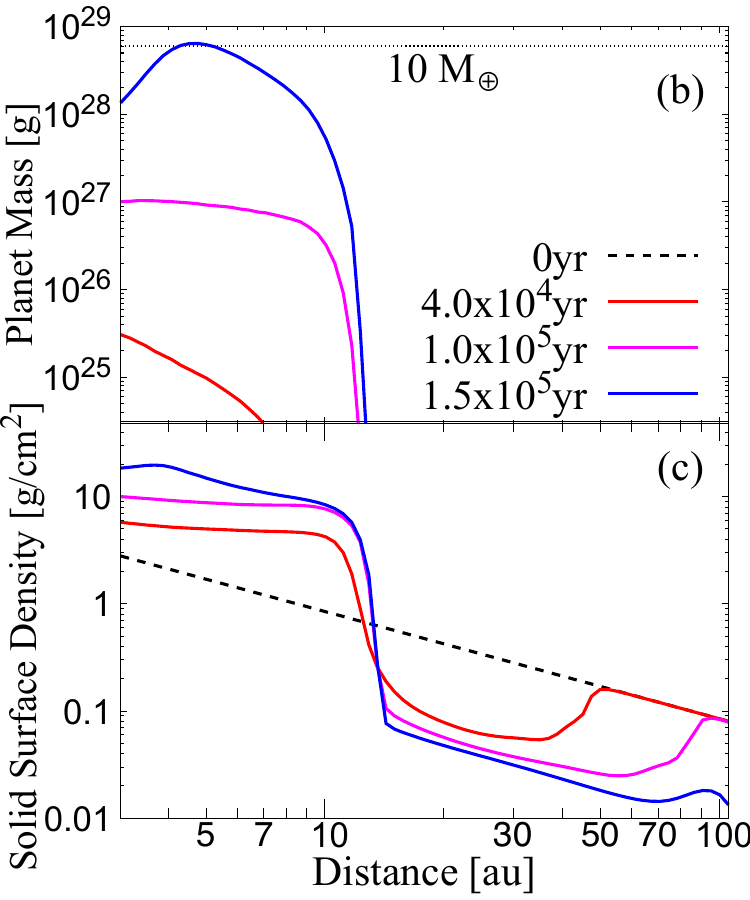}
\figcaption{Same as Figure~\ref{fig:dist_d1e-3}, but for $\rho_{\rm m} = 0.1\,{\rm g/cm}^3$. (a) is at $1.5 \times 10^5$ years. 
}
\label{fig:dist_d1e-1}
\end{figure*}

Pebbles are not formed yet beyond 60\,au (100\,au) at $7.0 \times 10^4$year ($1.6 \times 10^5$ years) so that the surface densities are similar to the initial ones (Figure~\ref{fig:dist_d1e-3}c). The formation of pebbles occurs at 40\,--\,60\,au (70\,--\,100\,au) at $7.0 \times 10^4$year ($1.6 \times 10^5$ years), which are called the growth fronts \citep{ohashi21}. The formation and drift of pebbles decreases the surface density inside the growth fronts. However pebbles grow into planetesimals in the inner disk ($\la 10$\,au), where the solid surface density is increased. The solid surface density enhancement induces the rapid formation of a massive core in the inner disk (Figure~\ref{fig:dist_d1e-3}b). It should be noted that the accretion of small planetesimals onto protoplanets contributes most to the rapid growth of protoplanets as mentioned in 
\citetalias{kobayashi21}. 

The result shown in Figure \ref{fig:dist_d1e-3} is slightly different from that in 
\citetalias{kobayashi21}, 
which is caused by the temperature difference. 
We here assume lower disk temperatures, resulting in slower drift velocities. 
Once the drifting pebbles with ${\mathrm St} \sim 1$ are much larger than the mean free path, they grow into planetesimals without significant drift. This increases the solid surface density around $\sim 10$\,au. However, the solid enhancement effectively occurs at 5\,--\,7\,au in 
\citetalias{kobayashi21}. 
Therefore, Figure~\ref{fig:dist_d1e-3} shows core formation slightly at the outer disk. In addition, core formation requires a longer time in the outer disk. The core formation timescale is slightly longer than that in 
\citetalias{kobayashi21}. 

We carry out the simulation with $\rho_{\rm m} = 0.1\,{\rm g/cm}^3$. Figure~\ref{fig:dist_d1e-1}a shows the mass distribution of bodies at $1.5 \times 10^5$\,years. In the inner disk with $r \approx 10$\,au, planetesimals with $m \ga 10^{15}$\,g are formed via collisional evolution, which is achieved if $\tau_{\rm grow} / \tau_{\rm drift} \la 0.7$ (see Equation~\ref{eq:timescale_ratio}). The runaway growth of planetesimals occurs at similar masses ($\sim 10^{17}$\,--\,$10^{19}$\,g) to the case of low $\rho_{\rm m}$. However, protplanets as massive as $10\,M_\oplus$ are formed at $\approx 5$\,au, which is smaller than the core formation location for $\rho_{\rm m} = 10^{-3} \,{\rm g/cm}^3$. In the outer disk, pebbles drift to the inner disk. The typical mass of pebbles are much smaller than the case with $\rho_{\rm m} = 10^{-3}\,{\rm g/cm}^3$ (cf. Figures \ref{fig:dist_d1e-3}a and \ref{fig:dist_d1e-1}a). The mass of pebbles are determined by ${\mathrm St}\sim 1$ so that their masses are small for high $\rho_{\rm b}$. 

Figure~\ref{fig:dist_d1e-1}b,c shows the protoplanet mass and the total solid surface density for $\rho_{\rm m} = 0.1\,{\rm g/cm}^3$. The massive cores for gas giants are formed at 4\,--\,5\,au at $1.5 \times 10^5$ years. Such the rapid formation of gas-giant cores is mainly caused by pebble drift and growth into planetesimals, which is the similar mechanisms to the fluffy pebble case ($\rho_{\rm m} = 10^{-3}\,{\rm g/cm}^3$). 

\begin{figure*}[htb]
\plottwo{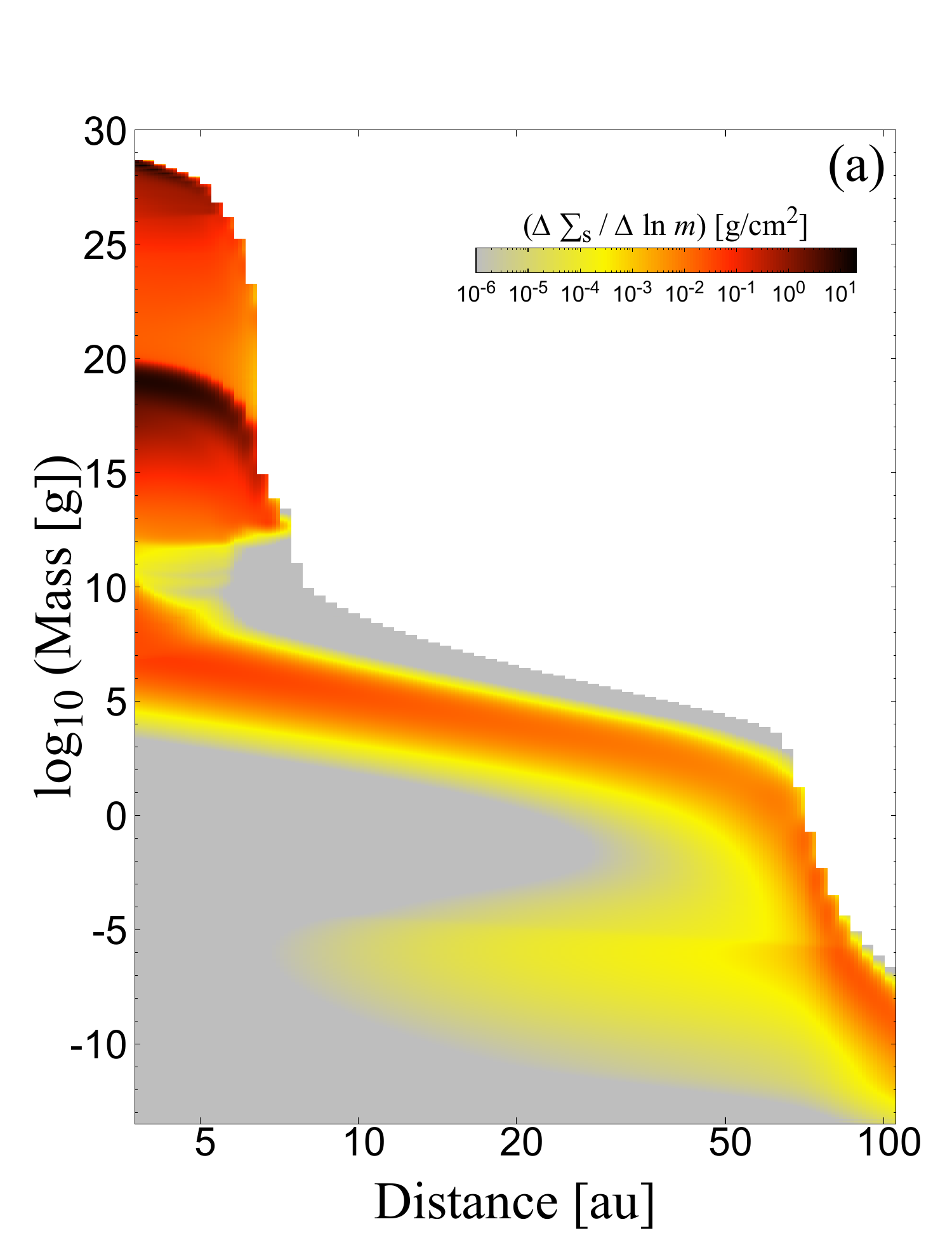}{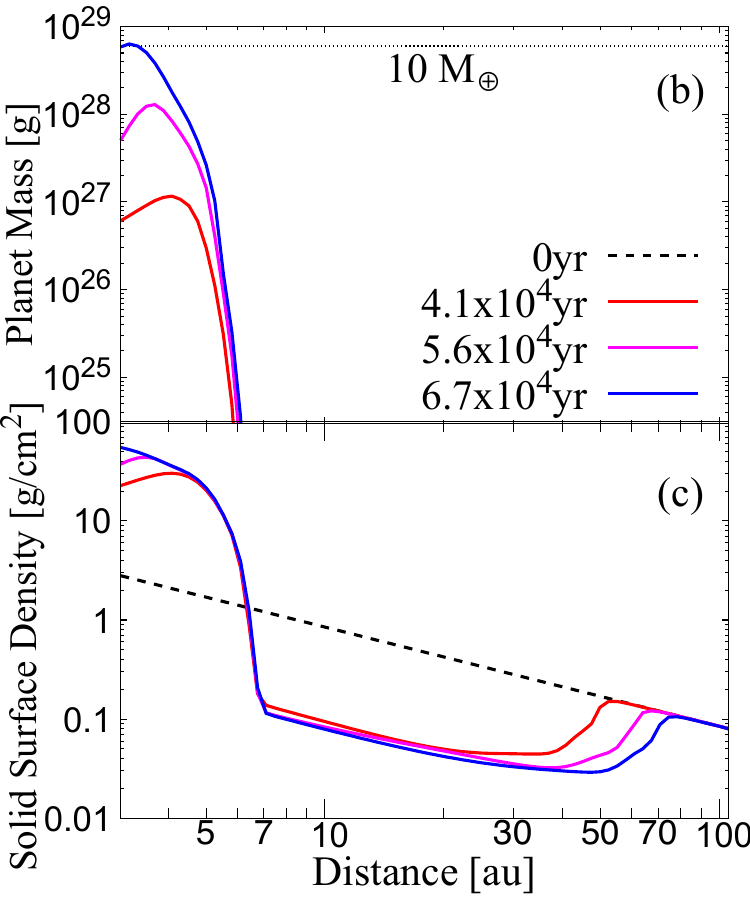}
\figcaption{Same as Figs.~\ref{fig:dist_d1e-3} and \ref{fig:dist_d1e-1}, but for $T_1 = 200$\,K and $\rho_{\rm m} = 0.1\,{\rm g/cm}^3$. (a) is at $6.7 \times 10^4$ years. }
\label{fig:dist_t200}
\end{figure*}

The radial profile of the surface density is different from the case for
fluffy pebbles ($\rho_{\rm m} = 10^{-3}\,{\rm g/cm}^3$). For fluffy
pebbles, pebbles drifting from the outer disk grow into planetesimals at
$r \approx 10$\,au, where the sizes of pebbles with ${\mathrm St} \sim
1$ is much larger than the mean free path. However, for non-porous
pebbles ($\rho_{\rm m} = 0.1\,{\rm g/cm}^3$), $\Sigma_{\rm s}$
similarly increases in $r \la 10$\,au, while the enhancement of
$\Sigma_{\rm s}$ is not so significant at the edge of the planetesimal
formation zone, $r \approx 10$\,au (Figure~\ref{fig:dist_d1e-1}c). 
Planetesimals are formed directly
from pebbles growing in the inner disk ($r \la 10$\,au) because $\tau_{\rm
grow}/\tau_{\rm drift} \la 0.7$. However, $\Sigma_{\rm s}$ of pebbles
drifting from the outer disk tends is smaller than that originally in
the inner disk. Pebbles drifting from the outer disk cannot achieve
the growth condition unless they drift to the more inner disk. Therefore, 
the enhancement of $\Sigma_{\rm s}$ more significantly occurs inside 5\,au. 
It should be noted that
pebbles can grow via collisions with previously forming planetesimals in
the inner disk. The growth condition is thus different from the estimate
of Equation~(\ref{eq:growth_time}) with the surface density of drifting
pebbles, because the estimate is derived under the assumption of the
single size population.


\subsection{Dependence on Temperature}

We perform the simulations for high disk temperatures with $T_1 = 200$ and 300\,K. The snowline of the disk with $T_1 = 300$\,K is at $\approx 4\,$au. \rev{We focus on the disk beyond 4\,au where the viscous heating is unimportant for disk temperatures as mentioned in \S~\ref{sc:disk_model}. }

Figure~\ref{fig:dist_t200}a shows the mass distribution of bodies at $6.7 \times 10^4$ years for $T_1 = 200$\,K and $\rho_{\rm m} = 0.1\,{\rm g/cm}^3$. Planetesimals with $m \ga 10^{15}$\,g are formed at $r \la 6$\,au, where $\tau_{\rm grow}/\tau_{\rm drift} \la 0.8$ (Equation~\ref{eq:timescale_ratio}). The onset of runaway growth determines the typical masses of planetesimals at $m \sim 10^{19}$\,g. Protoplanets with $\approx 10 M_\oplus$ are formed at 3\,--\,4\,au. Pebbles with $m \sim 10^4$\,--\,$10^9$\,g drift into the inner disk from the outer disk beyond 6\,au. 

\begin{figure}[htb]
\plotone{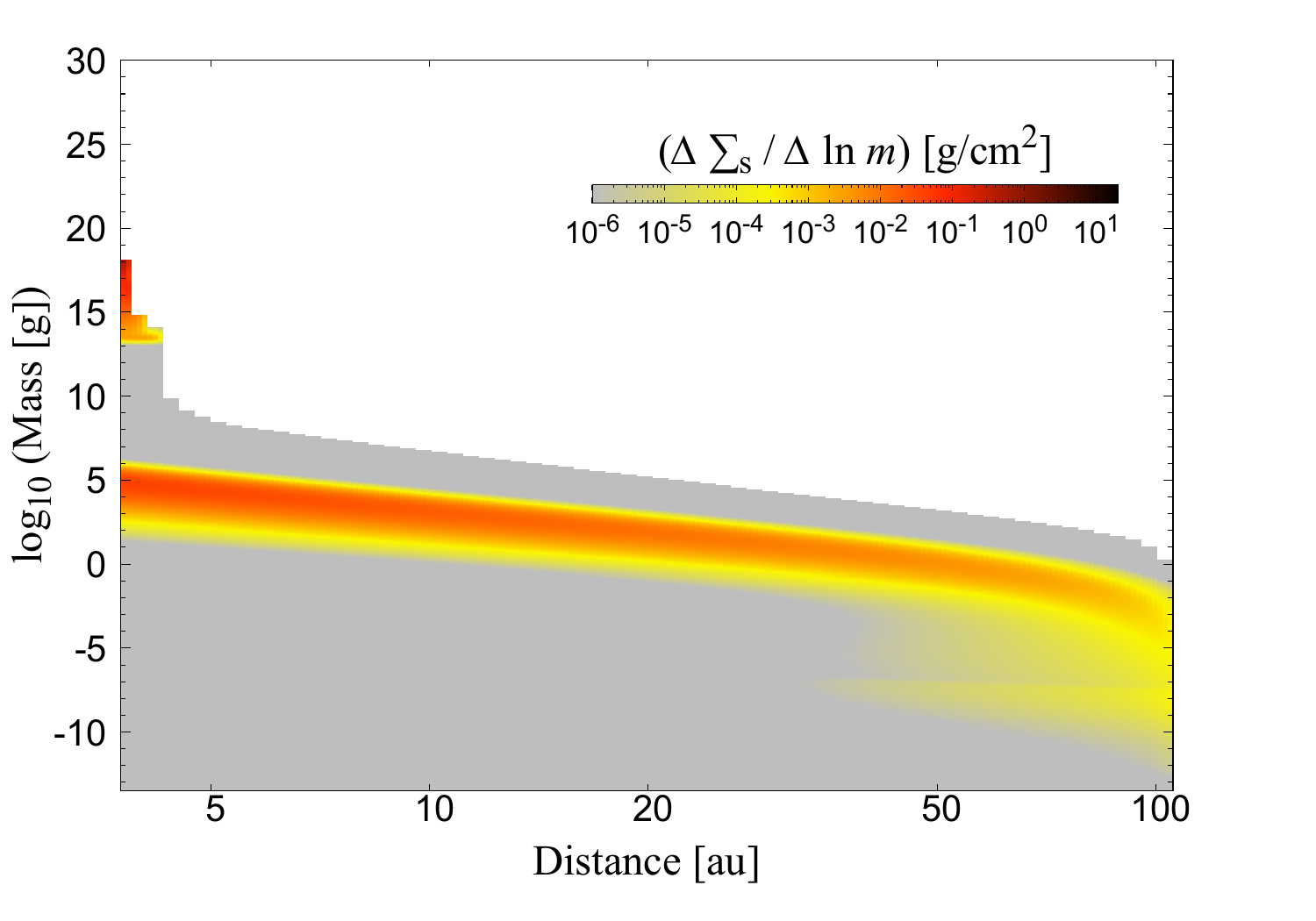}
\figcaption{Same as Figs.~\ref{fig:dist_d1e-3}a, \ref{fig:dist_d1e-1}a and \ref{fig:dist_t200}a, but at $1.4 \times 10^5$ years for $T_1 = 300$\,K and $\rho_{\rm m} = 0.1\,{\rm g/cm}^3$. 
The snowline is at $\approx 4\,$au.
}
\label{fig:dist_t300}
\end{figure}

Figure \ref{fig:dist_t200}b,c shows the protoplanet mass and the solid surface density for $T_1 = 200$\,K. Protoplanets with $10M_\oplus$ are formed at 3\,--\,4\,au at $6.7 \times 10^4\,$years (Figure~\ref{fig:dist_t200}b), which is caused by the solid surface density enhancement inside 6\,au (Figure~\ref{fig:dist_t200}c). According to the propagation of the growth front of pebbles, the solid surface density decreases inside the growth front beyond 6\,au, while the solid surface density increases inside 6\,au due to the growth of drifting pebbles into planetesimals. 
Higher disk temperatures induce planetesimal formation in the more compact inner disk. 
The solid surface density enhancement is more significant because pebbles are accumulated in the narrow planet forming region. In addition, the growth timescale is proportional to the Keplerian timescale, which is shorter for the inner disk. 
Therefore, massive cores are formed in a shorter timescale. 


However, planetesimals are not formed beyond the snowline in a higher temperature disk with $T_1 = 300$\,K (Figure~\ref{fig:dist_t300}). Thus gas giant cores are not formed beyond the snowline. Therefore, non-porous pebbles can form gas giant cores around 5\,au for $T_1 \la 200$\,K. This formation condition is achieved if $L_\star \la 11\,L_\sun$ for $M_\star = M_\sun$ (see Equation~\ref{eq:t1}). 
For a protostar with $M_\star \approx M_\sun$, the stellar luminosity is lower than $10\,L_\sun$ \citep{kunitomo17,stahler04}. Therefore, solar-type stars are likely to have gas giants if solid masses in disks are larger than $\approx 200\,M_\oplus$. The gas giant occurrence rate around solar type stars may be directly related to the distribution of solid masses in disks.

\section{Summery \rev{and Discussion}}
\label{sc:discussion}

We have investigated the growth from dust to gas-giant cores via ``dust-to-planet'' simulations (DTPSs). As shown in 
\citetalias{kobayashi21}, 
if pebbles are fluffy with bulk densities $\rho_{\rm b} \la 10^{-3}\,{\rm g/cm}^3$, collisional coagulation produces gas-giant forming massive cores at 5\,--\,10\,au in several $10^5$\,years (see also Figure~\ref{fig:dist_d1e-3}). However, large dust grains with such low bulk densities cannot explain the polarized millimeter wavelength lights from protoplanetary disks \citep{tazaki19}. Therefore, we have carried out DTPSs with higher densities of $\rho_{\rm b} \sim 0.1 \,{\rm g/cm}^3$. Gas giant cores are formed at $\approx 5$\,au in $\sim 10^5$\, years
for the nominal disk temperature with $T_1 = 100$\,K (see Equation~\ref{eq:temperature}), because of the growth of pebbles into planetesimals in the inner disk $\la 10$\,au (see Figure~\ref{fig:dist_d1e-1}). Therefore such rapid formation of gas-giant cores occurs even from non-porous pebbles. 

\rev{ Our simulations show collisional growth is limited by radial drift
in the outer disks
(Figures~\ref{fig:dist_d1e-3}--\ref{fig:dist_t300}). The size of bodies
basically agrees with the drift limit size estimated by
\citet{birnstiel12}. If collisional fragmentation occurs at velocities
larger than 1\,--\,10\,m/s, collisional fragmentation limits to small
pebbles.  However, even if planetary embryos are formed from such small
pebbles, pebble accretion onto embryos is inefficient for small pebbles
because of the gas flows around protoplanet atmospheres
\citep{okamura21}. Formation of gas-giant cores is not expected via
accretion of small pebbles. The state-of-art simulations for dust
aggregate collisions show collisional fragmentation occurs at much
higher velocities \citep{hasegawa23}. As discussed in
\S.~\ref{sc:DTPSmodel} and 
\citetalias{kobayashi21}, 
collisional fragmentation is
negligible. Especially fragmentaion hardly occurs in disks with low
temperature because of relatively low drift speeds of pebbles.  Pebbles
then grow into planetesimals in inner disks. DTPSs have consistently
performed the formation of planetary embryos and their growth to
giant-planet cores via planetesimal and pebble accretion. As analyzed in
\citetalias{kobayashi21}, 
the accretion rates of embryos are mainly determined by
planetesimal accretion, which is consistent with the classical theory
planetesimal accretion. However, the surface density of planetesimals
increases due to pebble growth into planetesimals and new born
planetesimals are relatively dynamically cold. The rapid growth of
giant-planet cores is achieved }
\citepalias[see][for details]{kobayashi21}. 

The growth of pebbles into planetesimals depends on the drift timescale so that the disk temperature is important (see \S \ref{sc:drift_growth}). 
Pebbles drifting from outer disks grow into planetesimals in lower temperature disks, resulting in gas giant core formation beyond 3\,au for $T_1 > 200$\,K (see Figure~\ref{fig:dist_d1e-1}--\ref{fig:dist_t300}). 
The disk temperature in an optically thick disk becomes high for high stellar 
luminosities. 
Although young stars tend to be luminous, 
the stellar luminosity for a star with $M_\star \approx M_\sun$ is lower than $10\,L_\sun$ even at protostar eras \citep{kunitomo17,stahler04}. 
Thus $T_1$ is estimated to be lower than 200\,K according to Equation~(\ref{eq:t1}). 
Therefore, gas giant cores are formed around such young luminous stars. 

\begin{acknowledgements}
We thank the reviewer for useful comments. 
The work is supported by
Grants-in-Aid for Scientific Research
(21K03642, 22H01278, 22H01286, 22H00179) 
from MEXT of Japan. 
\end{acknowledgements}

\bibliography{apj}

\end{document}